\documentclass[reprint, amsmath, amssymb, aps]{revtex4-2}

\usepackage{graphicx} 
\usepackage[utf8]{inputenc}
\usepackage{amsmath}
\usepackage{ulem}
\usepackage{hyperref}

\usepackage{xcolor}

\begin{document}
\title{Regression on Regression: Mapping Data-Driven Binary Black Hole Merger Rate Fits to Progenitor Histories}

\author{Emma Blanchet}
\affiliation{Canadian Institute for Theoretical Astrophysics, 60 St George St, University of Toronto, Toronto, ON M5S 3H8, Canada}
\email{emma.blanchet@outlook.com}

\author{Aryanna Schiebelbein-Zwack}
\affiliation{Canadian Institute for Theoretical Astrophysics, 60 St George St, University of Toronto, Toronto, ON M5S 3H8, Canada}
\affiliation{David A. Dunlap Department of Astronomy and Astrophysics, University of Toronto, 50 St George St, Toronto ON M5S 3H4, Canada}

\author{Maya Fishbach}
\affiliation{Canadian Institute for Theoretical Astrophysics, 60 St George St, University of Toronto, Toronto, ON M5S 3H8, Canada}
\affiliation{David A. Dunlap Department of Astronomy and Astrophysics, University of Toronto, 50 St George St, Toronto ON M5S 3H4, Canada}
\affiliation{Department of Physics, 60 St George St, University of Toronto, Toronto, ON M5S 3H8, Canada}

\begin{abstract}

The binary black hole (BBH) merger rate is governed by the progenitor formation rate and the distribution of delay-times between formation and merger, but these functions remain poorly constrained. We introduce a framework that maps the parameters of physics-driven models directly onto existing data-driven fits of the BBH merger rate. This ``regression on regression” approach enables physical interpretation of flexible population models without the computational burden of reanalyzing the underlying gravitational-wave event data. Applying this method to the \textsc{B-Spline} merger-rate posteriors from the Fourth Gravitational-Wave Transient Catalog, we fit the minimum delay time ($\tau_{\text{min}}$), delay-time power-law index ($\alpha$), and progenitor formation parameters controlling the normalization ($\mathcal{A}$), early-time growth ($\gamma$), and late-time decay ($\delta$). Increasing the number of anchoring redshift points from two to four reduces the median sum-squared error (SSE) by a factor of $\approx 4.5$. However, residuals reveal that the physical model does not pass through all four anchors, exposing model misspecification and demonstrating a key strength of the framework: unlike standard inference methods, which preferentially weight compatible curves and mask underlying tensions, our approach exposes BBH posteriors irreconcilable with the model. Despite uncertainties at $z\gtrsim1$, the shape of the progenitor formation rate at low-$z$ is robust and evolves more steeply than the global star formation rate (SFR), supporting a preference for low metallicity environments. Specifically, the log-space slope of the progenitor rate is $\approx 5.3$ times steeper than the SFR between $z=0.1$ and $z=1.0$. Ultimately, a more complex phenomenological model is required to match the \textsc{B-Spline} merger rates.
\end{abstract}

\maketitle

\section{Introduction}
\label{sec:intro}

Over the last decade since the first gravitational wave detection, the LIGO-Virgo-KAGRA~\citep[LVK;][]{2015CQGra..32g4001L,2015CQGra..32b4001A,2021PTEP.2021eA101A} collaboration's Gravitational-wave Transient Catalog (GWTC) has grown to over a hundred compact object mergers~\citep{2025arXiv250818082T}. These detections have enabled increasingly accurate studies of the population properties of compact objects, including the mass and spin distributions of binary black holes (BBHs), as well as their merger rate and its evolution across cosmic time. With current observations, the merger rate can be probed out to $z \approx 1.5$ \citep{LVKgwtc4}. This sensitivity traces black hole (BH) mergers back nearly eight billion years, reaching into the first half of cosmic history. 

With their latest catalog GWTC-4.0, the LVK fit the BBH merger rate as a function of redshift using both a strongly-modeled \textsc{Power-Law Redshift} parameterization, in which the merger rate evolves as $\mathcal{R}_\mathrm{BBH} \propto (1+z)^\kappa$~\citep{LVKgwtc4}, as well as a flexible, data-driven \textsc{B-Spline} model in which the power-law is modulated by basis splines~\citep{2023ApJ...946...16E}. While the \textsc{B-Spline} reconstruction is broadly consistent with the power-law parameterization, it favors additional structure in the merger rate evolution, including a flatter trend between $z=1.5$ and $z=1$, followed by a steeper decline toward $z=0$. Using the \textsc{Power-Law Redshift} model, LVK inferred a BBH merger rate at $z=0.2$ of $\mathcal{R}_\text{BBH} = 29^{+8.5}_{-6.5} \text{Gpc}^{-3} \text{yr}^{-1}$ ~\citep{LVKgwtc4}, whereas the \textsc{B-Spline} analysis yielded a higher value of $\mathcal{R}_\text{BBH} =38^{+19}_{-10} \text{Gpc}^{-3} \text{yr}^{-1}$ at the same redshift~\citep{LVKgwtc4}. Despite these differences, both models indicate that the BBH merger rate decreases by approximately an order of magnitude from $z=1.5$ to the present day.

While data-driven approaches such as the \textsc{B-Spline} model benefit from the flexibility to accurately describe features in the BBH population compared to rigid parametric models, they are consequently challenging to interpret. 
Physics-driven models, on the other hand, describe the population model in terms of physically meaningful, interpretable parameters, but they can be computationally expensive and suffer from model misspecification if the physical model is wrong or incomplete \citep{boesky_binary_2024, 2026arXiv260120202L}.
Previous studies developed methods to parametrically reconstruct data-driven population fits, thereby extracting physically meaningful parameters without refitting the data~\citep{2022mla..confE..25W,2025PhRvD.111j4053F,2024PhRvX..14b1005C,2025JCAP...12..031R,2026arXiv260420941C}. We build on these ideas by developing a method to directly translate data-driven fits of the BBH merger rate into an interpretable, physics-driven model; effectively performing a regression on the LVK regression itself. In this way, our approach preserves the flexibility of data-driven reconstructions while providing a direct mapping onto the underlying physical model parameters.

In order to interpret the redshift evolution of the BBH merger rate, we employ a physically motivated model that links BBH mergers to their formation processes. We assume that the BBH merger rate depends on their progenitors' formation rate convolved with a delay time distribution~\cite{Mapelli2021, Fishbach_kalogera,Fishbach_vonson2023,Fishbach_2025}:
\begin{equation}\label{eq:mergerrate}
\begin{split}
    \mathcal{R}_{\text{BBH}}(t) = \int^{\tau_{\text{max}}}_{\tau_{\text{min}}} \mathcal{R}_{\text{prog}}(t- \tau) p(\tau) d\tau.
\end{split}
\end{equation}

The rate of BBH mergers is related to the rate at which their progenitor stars form, $\mathcal{R}_{\text{prog}}$, but we must also account for the delay time, $\tau$, between when the progenitor stars form, and when they ultimately merge. The delay time distribution, $p(\tau)$, depends on the formation scenarios that lead to merging BBH systems, including binary stellar interactions and/or stellar dynamics~\citep[see, e.g.][for a review]{2020FrASS...7...38M,Mandel_2022}. These different formation conditions can produce varied initial separations between the BHs and degrees of orbital separation shrinkage over time, leading to varied inspiral timescales. The distribution of delay times is typically approximated as a power law,
\begin{equation}\label{eq:ptau}
\begin{split}
    p(\tau) = \frac{\alpha + 1}{\tau^{\alpha + 1}_{\text{max}} - \tau^{\alpha + 1}_{\text{min}}} \tau^{\alpha},
\end{split}
\end{equation}
for delay times, $\tau$, between $\tau_\mathrm{min}$ and $\tau_\mathrm{max}$. We fix the maximum delay time to the age of the Universe, $\tau_\mathrm{max} = 13.8$ Gyr, and fit for the minimum delay time $\tau_\mathrm{min}$ and the power law slope $\alpha$, which are both informative of formation mechanisms.
More generally, $\alpha$ reflects the characteristic merger timescale: pathways that efficiently produce short delays tend to yield steeper, more negative slopes (e.g., $\alpha \lesssim -1$), whereas channels that preferentially produce wider binaries and longer inspiral times give shallower distributions (e.g., $\alpha \approx -0.3$ to $-0.5$). Typical BBH formation scenarios predict power-law slopes in the approximate range 
$-2 < \alpha < -0.3$~\citep{Dominik2012,2017MNRAS.471.4702B,2022MNRAS.516.5737B,Rodriguez2019,Ye_2024,Fishbach_vonson2023,Gallegos_Garcia_2021,vanSon2022, 2023MNRAS.523.4539K}. Thus, constraints on delay time parameters from gravitational wave observations provide a useful diagnostic of the dominant formation history of the BBH population.

The BBH progenitor formation rate, $\mathcal{R}_\text{prog}$, is related to the star formation rate (SFR), accounting for the fact that the BBHs observed by the LVK collaboration likely originate from massive stellar progenitors.
While the formation rate of BBH progenitors is expected to deviate from the global SFR, because BBH progenitors are a biased subset of all stars, we generally expect it to follow a similar structure to the SFR: rising at early times before turning over and declining towards late times.
We therefore fit a functional form for the progenitor formation rate inspired by the shape of the SFR as parameterized by~\citet{Katsianis} with a power-law rise and exponential decay:
\begin{equation} \label{eq:prog}
    \mathcal{R}_{\text{prog}} (t) = \mathcal{A} t^{\gamma} e^{\delta t}\times 10^{-6} \text{Mpc}^{-3}\,\text{yr}^{-1},
\end{equation}
where $t$ is the age of the Universe given in Gyr, $\mathcal{A}$ controls the amplitude of the progenitor formation rate (normalized by $10^{-6}\,M_\odot^{-1}$ to account for the approximate efficiency of BBH formation \citep{Fishbach_vonson2023}), $\gamma$ represents the power law index for early time growth, and $\delta$ (in units of $\text{Gyr}^{-1}$) governs the exponential decline at late times.

The proportion of stellar mass that ends up in BBH mergers can be described using an efficiency factor $\epsilon$ (conventionally in units of $M_\odot^{-1}$):

\begin{equation}\label{eq:efficiency}
\begin{split}
    \epsilon = \frac{\mathcal{R}_{\text{prog}}}{\text{SFR}} .
\end{split}
\end{equation}

The efficiency depends on key elements of stellar composition and environment. An important property in determining a star’s evolution and final fate is its initial metallicity~\citep{chruslinska2022chemicalevolutionuniverseconsequences}. 
Stars born in metal-rich environments are expected to produce merging BBHs less efficiently than those forming in metal-poor conditions \citep{Gallegos_Garcia_2021, Mandel_2022, Mapelli2021, schiebelbeinzwack2024massdensitymergingbinary, Chruslinska2020, chruslinska2022chemicalevolutionuniverseconsequences,2025ApJ...979..209V, Belczynski2010, 2008NewAR..52..419V, 2002MNRAS.329..897H, 2008MNRAS.391.1117P}.
This is because high iron traces in outer stellar layers produce stronger stellar winds, leading to mass loss and greater radial expansion~\citep{Kudritzki_puls}.
This makes it less likely for iron-rich stars to retain sufficient mass to collapse to a BH. Additionally, their increased radius promotes earlier mergers with stellar companions, preventing BBH formation. 
Since the star-forming metallicity of the Universe increases with its age, $\epsilon$ is expected to decrease with time.
The progenitor formation rate of merging BBHs is therefore expected to peak earlier (at higher redshifts) than the global SFR and steeply decline towards low redshifts. 
Indeed, analyses using GWTC-3.0 favor a steep evolution of the progenitor formation rate coupled with short delay times characterized by power-law slopes $\alpha < -1$~\citep{abbot_et_al, schiebelbeinzwack2024massdensitymergingbinary,Fishbach_vonson2023,Fishbach_kalogera,2024ApJ...967..142T,Wu_2024}, a result that we validate here with GWTC-4.0.

In this work, we develop a new method to extract the delay time distribution and progenitor formation rate without refitting the BBH merger rate population, and apply our method to GWTC-4.0.
Our approach is similar to the method proposed by \citet{Fishbach_kalogera} of fitting the delay time distribution and formation rate parameters by using previous fits to the BBH merger rate inferred at two redshift points, $z = 0$ and $z = 1$. In their approach, a data-driven or phenomenological model is first fit to the gravitational wave data to estimate $\mathcal{R}_{\rm BBH}(z)$. Rather than using the posterior on the merger rate at all redshift points,  \citet{Fishbach_kalogera} isolate the joint posterior on $\mathcal{R}_{\rm BBH}(z = 0)$ and $\mathcal{R}_{\rm BBH}(z = 1)$, and use this to derive an approximate likelihood from which to sample the delay time distribution and formation rate parameters of interest. This method was later applied to GWTC-3.0 by Refs.~\citep{2023MNRAS.522.5546F,2024ApJ...972..157V,schiebelbeinzwack2024massdensitymergingbinary,Wu_2024}{, albeit with varied details on the definition of the progenitor formation rate. For example, \citet{schiebelbeinzwack2024massdensitymergingbinary} explicitly model the BBH formation efficiency as a function of the redshift-evolving cosmic metallicity, whereas in this work we define the efficiency with Eq. \ref{eq:efficiency}, using our derived $\mathcal{R}_\text{prog}$ as a proxy for metallicity.}
A limitation of this method, however, is that it is computationally expensive to use more than two redshift points along the inferred $\mathcal{R}_{\rm BBH}(z)$.
This means that if there are interesting features in the BBH merger rate evolution at intermediate redshifts, they cannot be captured with this approach.
We therefore develop a new method based on simple least squares regression that fits an arbitrary $\mathcal{R}_{\rm BBH}(z)$ curve drawn from the gravitational wave posterior in terms of the physical model of interest (in our case, Eq.~\ref{eq:mergerrate}). In our method, it is straightforward to use as many redshift points on the $\mathcal{R}_{\rm BBH}(z)$ posterior curves as desired, thus capturing nontrivial structure in the merger rate evolution as suggested, for example, by the \textsc{B-Spline} data-driven population results. Another advantage of this approach is that it enables a more comprehensive exploration of the model's ability to match with all of the merger rate curves, whereas traditional likelihood-based methods ({such as} those in {e.g.} \citep{Fishbach_kalogera, schiebelbeinzwack2024massdensitymergingbinary, Wu_2024}) only latch onto curves it can describe well, while ignoring the others. Accounting for each posterior draw of the merger rate is essential, as neglecting specific draws risks overlooking instances where the data-driven results are fundamentally at odds with certain formation models \citep{SZ2026}. {More broadly, flexible population reconstructions such as the LVK \textsc{B-Spline} merger rate provide low-bias descriptions of the data at the expense of interpretability, whereas phenomenological models offer more interpretable but higher-bias representations of the merger rate evolution. Our framework provides an interpretable projection of these nonparametric posteriors onto physically motivated models, allowing any mismatch to diagnose model bias.}

The remainder of this paper is structured as follows. In Section \ref{sec:methods}, we detail our least squares regression framework for mapping flexible, data-driven merger rate reconstructions onto the parameter space of physical progenitor models. We present our primary findings in Section \ref{sec:results}, beginning with a comparison of model adherence when using two anchoring redshift points (Section \ref{sec:results_2point}) versus a four-point implementation (Section \ref{sec:results_4point}). Subsequently, we describe the progenitor formation rates and delay time distributions inferred from the four-point curve-fit in Section \ref{sec:results_progform}. Finally, we summarize our conclusions in Section \ref{sec:conclusion}.

\section{Methods}\label{sec:methods}

We infer the BBH delay time distribution and progenitor formation rate from the latest LVK catalog, GWTC-4.0.
Rather than directly fitting the BBH merger rate model of Eq.~\ref{eq:mergerrate} to GWTC-4.0 using traditional hierarchical Bayesian population inference, we start with a data-driven fit to $\mathcal{R}_{\rm BBH}(z)$ from the \textsc{B-Spline} model~\citep{2023ApJ...946...16E,LVKgwtc4, ligo_scientific_collaboration_2025_16911563}. The posterior draws from the \textsc{B-Spline} fit are publicly available at \url{https://zenodo.org/records/16911563}. 

We use the well-known technique of least squares regression, as implemented in SciPy’s \texttt{curve\_fit} function, to fit Eq.~\ref{eq:mergerrate} to the \textsc{B-Spline} posterior draws \citep{Virtanen2020SciPy}. Specifically, we minimize the sum-squared difference:
\begin{equation}
f(\Lambda) = \sum_i\left(\log\mathcal{R}_{\rm BBH}^{\rm model}(z_i) - \log\mathcal{R}_{\rm BBH}^{\rm data}(z_i)\right)^2,
\end{equation}
where $\Lambda$ are the delay time distribution (Eq.~\ref{eq:ptau}) and progenitor formation rate (Eq.~\ref{eq:prog}) parameters and we choose fixed redshift points $z_i$. 

To constrain the fit, we specify boundary values for the parameters that make up $\Lambda$:
 (i) $\tau_{\rm min}/{\rm Gyr} \in (0.01, 4)$, (ii) $\alpha \in (-3, -0.001)$, (iii) $\log(\mathcal{A}) \in (-8, 4)$, (iv) $\gamma \in (-8, 5)$, and (v) $\delta \times \mathrm{Gyr^{-1}} \in (-8, -10^{-5})$.
These ranges encompass the necessary coverage to fit the data, as seen in Figure \ref{fig:cornerplot_2p} and Figure \ref{fig:cornerplot_4p}, except in instances where we observe boundary-railing behavior or degeneracies. These behaviors effectively expose where the physics-driven model is unable to properly match particular merger rate posteriors, which we discuss in the following sections. 

For the choice of redshift points $z_i$, we consider two cases: using the redshift points $z=0.2$ and $z=1.0$, matching the previous approach introduced by~\citet{Fishbach_kalogera} that used only two redshift points, and our improved scenario that uses four redshift points $z=0$, $z=0.4$, $z=1.0$ and $z=1.5$, and thus includes more detailed structure in the merger rate evolution. In the first case, the 2-point framework focuses on $z=0.2$ and $z=1.0$ instead of the boundary points ($z=0$ and $z = 1.5$) because the \textsc{B-spline} model agrees better with the parametric \textsc{Power-law redshift} model at these points, possessing smaller uncertainties ~\cite{LVKgwtc4}.
In the second case, we restrict our analysis to only these four points in order to focus on the overall merger rate shape rather than the detailed structure of the splines, whose numerous smaller scale fluctuations are unlikely to be physically meaningful~\citep{2023ApJ...955..107F, SZ2026}. {We choose $z=0.4$ specifically because previous work has suggested potential structure in the merger rate evolution near this redshift \citep{2024PhRvX..14b1005C}. We have additionally verified that the results presented in Section~\ref{sec:results_4point} are not affected by changes in the choice of anchor points, indicating that our conclusions are robust to the specific selection of the four redshift points.}

We iterate over every $\mathcal{R}_\text{BBH}$ curve in the \textsc{B-Spline} posterior to compute the best fits; i.e., those that minimize $f(\Lambda)$. For each posterior draw, the resulting minimum sum-squared error (SSE), i.e.,
\begin{equation}
\label{eq:SSE}
    {\rm SSE} = {\rm min}_\Lambda(f(\Lambda))
\end{equation}
provides a goodness-of-fit measure of how closely the model can reproduce the given data-driven posterior draw.
{This curve-fitting technique maps the data-driven posterior on $\mathcal{R}_{\rm BBH}(z)$ to point estimates of $\Lambda$, effectively building an empirical distribution of the parameters governing the BBH delay time distribution and progenitor formation rate.}

Furthermore, the distribution of SSE values across the posterior draws measures how well the physics-driven model can reproduce the features in the data-driven model.
\begin{figure}
    \centering
   {\includegraphics[width=1\linewidth]{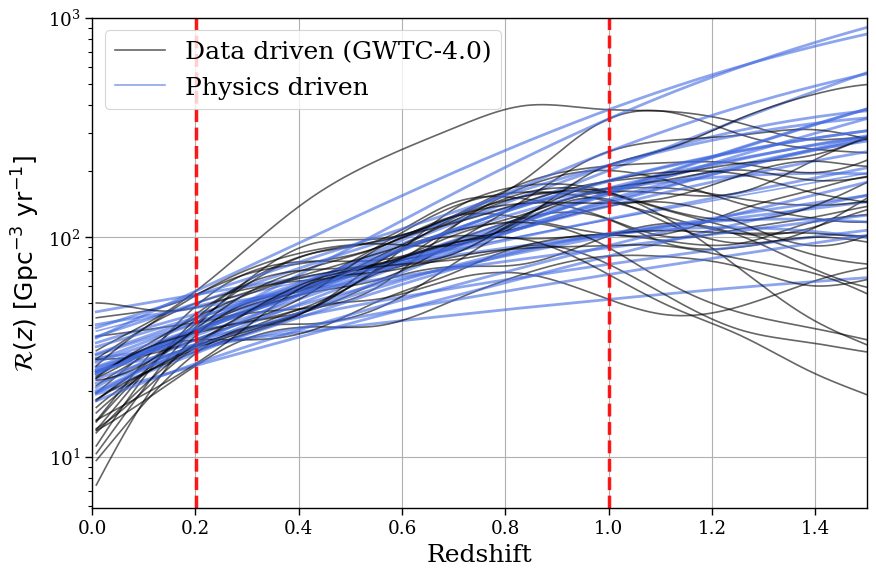} }%
    \qquad
    {\includegraphics[width=1\linewidth]{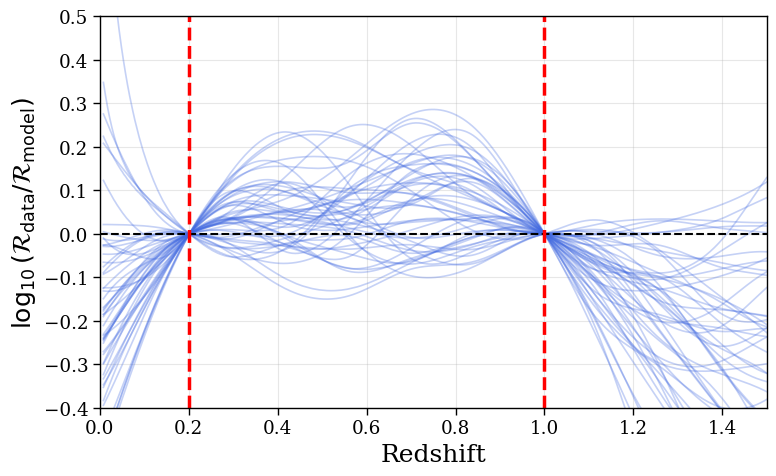} }%
    \caption{\textit{Top:} BBH merger rate as a function of redshift. A subset of \textsc{B-Spline} posterior draws (black) overplotted with their corresponding physics-driven model fit (blue) for the 2-point curve-fitting optimization. \textit{Bottom:} Residuals between data-driven GWTC-4.0 posterior draws and the corresponding best-fit physics-driven model for the 2-point fit. The red dashed vertical lines show the redshift points used, $z=0.2$ and $z=1.0$. 
    }
    \label{fig:plot_2point}
\end{figure}

\begin{figure}
    \centering
   {\includegraphics[width=1\linewidth]{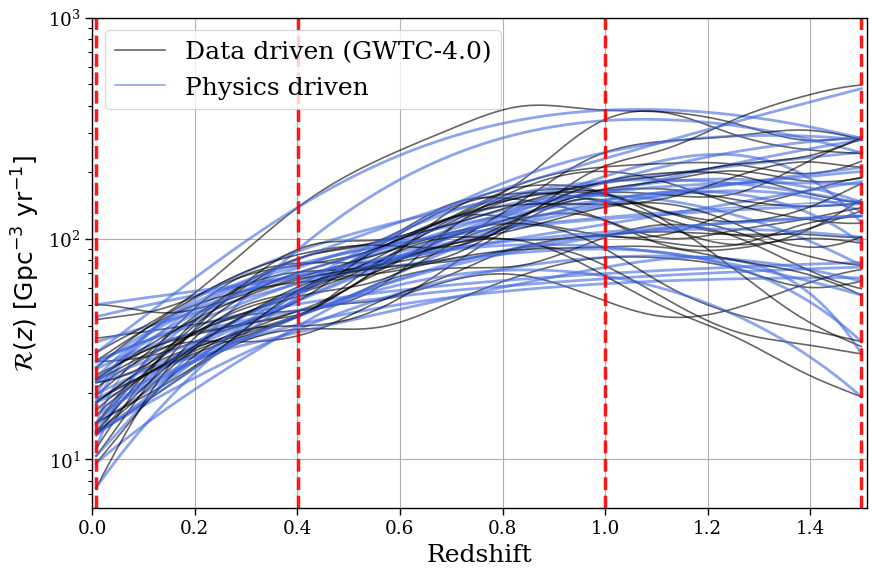} }%
    \qquad
    {\includegraphics[width=1\linewidth]{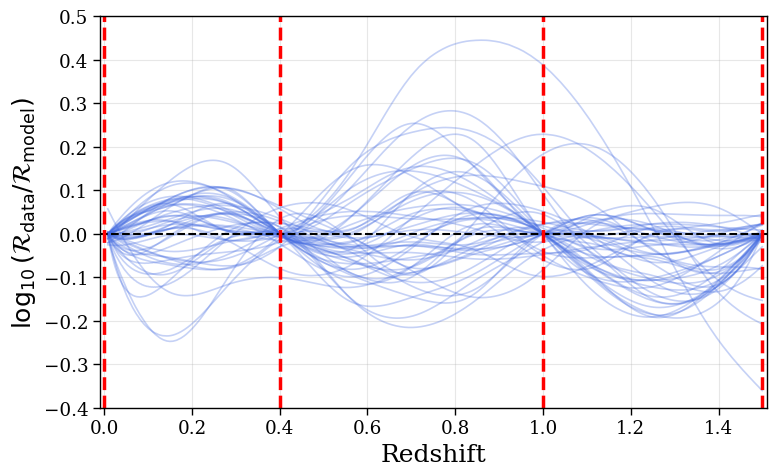} }%
    \caption{\textit{Top:} BBH merger rate as a function of redshift. {The subset of \textsc{B-Spline} posterior draws (black) is the same as those in the top panel of Figure \ref{fig:plot_2point}}. The 4-point curve fitting provides a better fit to the data, as seen visually {(in blue)}, as well as confirmed quantitatively computed via the SSE values. \textit{Bottom:} residuals between data-driven GWTC-4.0 posterior draws and the corresponding best-fit physics-driven model for the 4-point curve-fitting. The physics-driven model better reproduces the structure in the data-driven posterior compared to the 2-point fit, but the residuals are still large at the intermediate redshift points. The redshift points are shown in red at $z=0$, $z=0.4$, $z=1.0$ and $z=1.5$.
    }
    \label{fig:plot_4point}
\end{figure}
\section{Results \& Discussion} \label{sec:results}

We now apply the 2-point and 4-point curve-fitting framework introduced in the previous section to the GWTC-4.0 \textsc{B-Spline} results. We present the inferred delay time distribution and progenitor formation rate parameters, and compare our physically motivated fit to the original \textsc{B-Spline} merger rate inference. 

We highlight that the 4-point fit is better able to reproduce the global structure in the BBH merger rate evolution identified by the \textsc{B-Spline} fit in Section \ref{sec:results_4point}. 
Using our 4-point results, we focus on our inferred progenitor formation rate and derive the BBH progenitor formation efficiency as a function of redshift according to Eq.~\ref{eq:efficiency} in Section \ref{sec:results_progform}.

\subsection{2-point Fit}\label{sec:results_2point}

In the 2-point framework, we use the BBH merger rate measured at only two $z$ points, $z=0.2$ and $z=1.0$, to constrain the delay time distribution and progenitor formation rate parameters.
As seen in the top panel of Figure \ref{fig:plot_2point}, the model's best-fitting merger rates reproduce the amplitudes at both specified $z$ points, but fail to capture the overall shape of the data-driven draws across redshift, with significant deviations between the model and \textsc{B-Spline} results in the intermediate redshift range and beyond $z>1$. Calculating the SSE for each posterior draw (Eq.~\ref{eq:SSE}), we find $\text{SSE}=13.38 ^{+53.55}_{-2.37}$ at 90\% credibility. This distribution is heavily right-skewed, reflecting an incompatibility between the physics-driven model and a portion of the BBH merger rate posterior draws. This aligns with \citet{SZ2026}, who noted that there are fundamental physical incompatibilities of particular merger rate posteriors with conventional delay time distributions. Our curve-fitting method provides another way to quantify this tension; the high-SSE tail effectively isolates the specific merger rate realizations that cannot be explained by the physics-driven model. {As we explore further in Section \ref{sec:results_4point}, the $\mathcal{R}_\text{BBH}$ curves yielding the largest SSE values correspond to realizations that exhibit a pronounced peak within $z < 1.5$. The physics-driven model, constrained by its functional form, cannot reconcile this turnover with the necessary growth parameters, resulting in high SSEs.}

In addition to the SSE values, to better examine the goodness of fit, we compute the residuals between the GWTC-4.0 curves and their fitted physics driven model curves as seen in the bottom panel of Figure \ref{fig:plot_2point}. The median residual distributions at redshift $z=0.2$ and $z=1.0$ are tightly constrained around 0, with 90\% credible intervals of $\log{\frac{\mathcal{R}_\text{data}}{\mathcal{R}_\text{model}}}=0^{+1.2}_{-1.1} \times 10^{-8}$ and $\log{\frac{\mathcal{R}_\text{data}}{\mathcal{R}_\text{model}}}=0^{+1.1}_{-1.2} \times 10^{-8}$ respectively. Visually, the discrepancy between the data and the model curves is greater near the boundaries under the 2-point framework (as seen in the bottom panel of Figure \ref{fig:plot_2point}). This suggests that while the model is able to find suitable parameters to match the rate at redshifts $z=0.2$ and $z=1.0$, it ignores the rest of the curve structure. Specifically, the model does not capture the flattening occuring in the merger rate for $z>1$ with the model systematically overestimating $\mathcal{R}_\text{BBH}(z=1.5)$.

\begin{figure*}
    \centering
    \includegraphics[width=10cm]{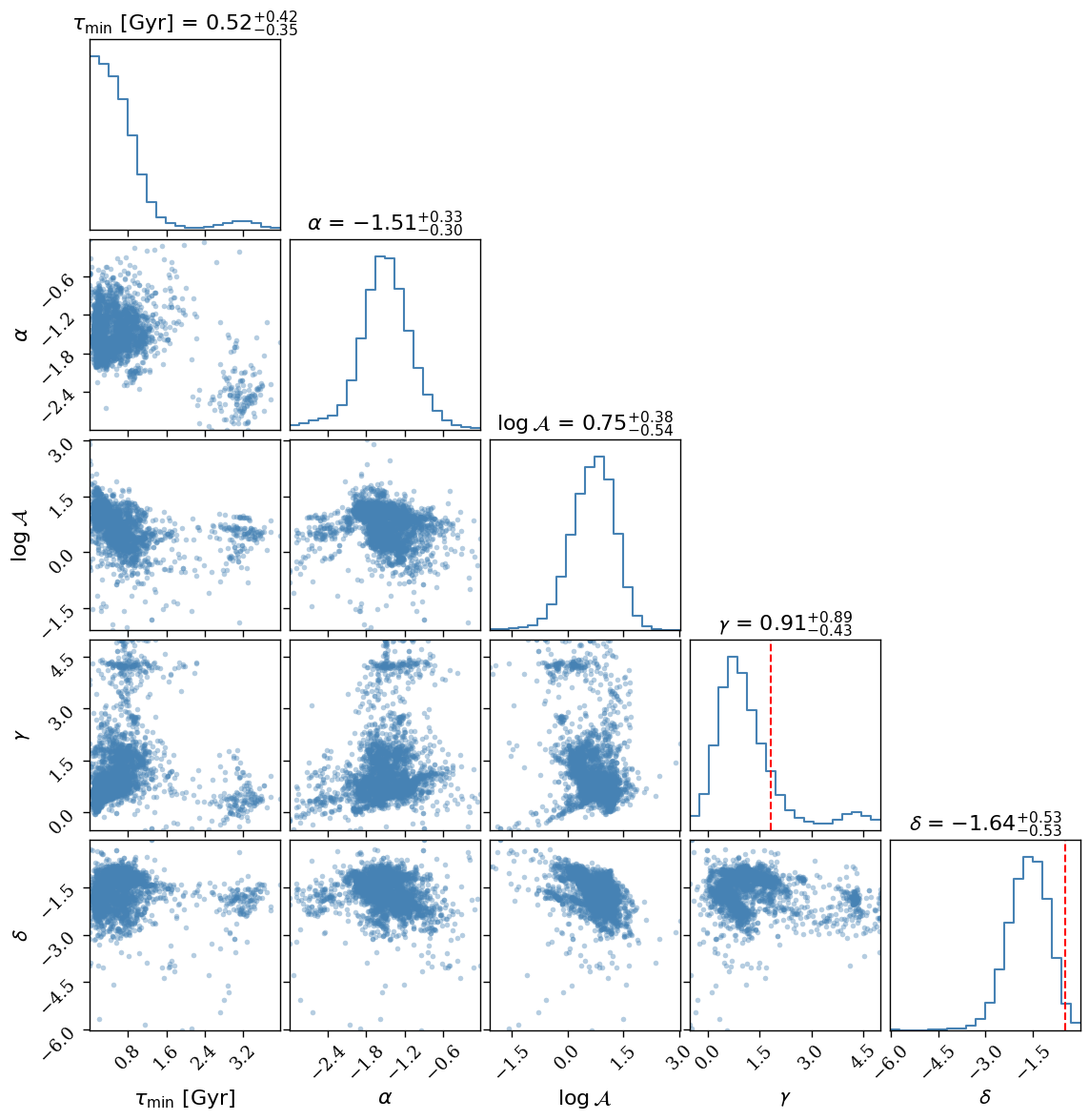}
    \caption{The inferred parameter distributions from the 2-point optimization approach at $z=0.2$ and $z=1.0$. The delay time parameters are $\tau_\text{min}$ and the power law index $\alpha$. The progenitor formation rate parameters are the normalization $\mathcal{A}$, the power law growth at early cosmic times $\gamma$, and the exponential decline at late cosmic times $\delta$. Each scatter point represents the curve-fit to a single BBH merger rate posterior draw. Red vertical lines indicate the values for the global SFR \citep{Katsianis}.}
    \label{fig:cornerplot_2p}

    \centering
    \includegraphics[width=10cm]{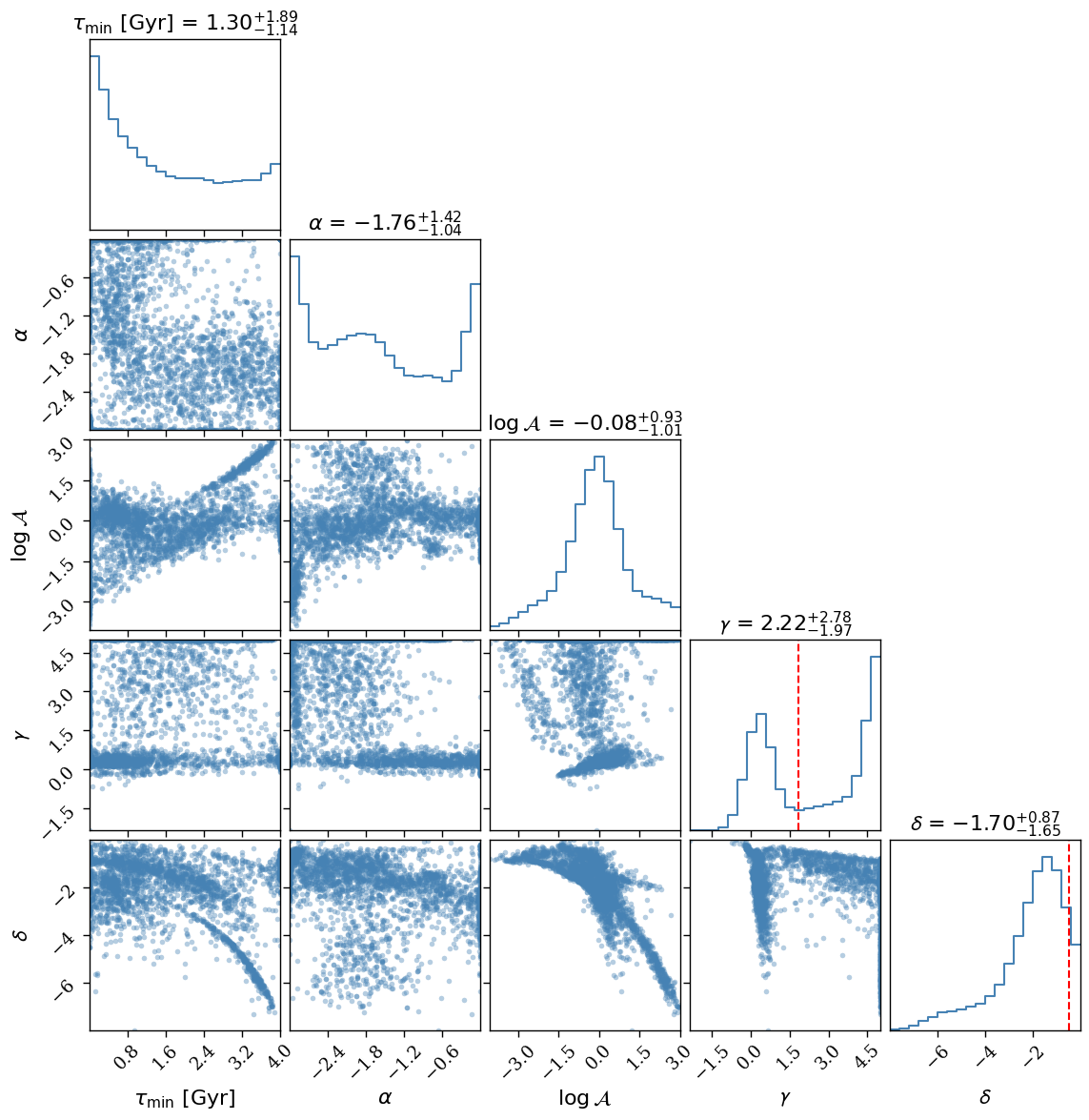}
    \caption{Similar to Fig. \ref{fig:cornerplot_2p}, but using the 4-point approach at $z=0$, $z=0.4$, $z=0.6$ and $z=1.5$. The additional points require the model to replicate more of the structure in the data. This can lead to degeneracies and railing behaviour due to exposed model misspecification.}
    \label{fig:cornerplot_4p}
\end{figure*}

The best-fit delay time and progenitor formation rate parameters corresponding to each \textsc{B-Spline} posterior draw are shown in Figure \ref{fig:cornerplot_2p}. 
We find a preference for a relatively steep delay time distribution with a power law slope $\alpha = -1.51^{+0.33}_{-0.30}$ at 90\% credibility, broadly consistent with the range of population synthesis simulations, albeit steeper than the typical prediction of $\alpha \approx -1$ for, e.g., common envelope, which is considered an efficient merger mechanism. 
Meanwhile, our fit also favors smaller values for the minimum delay time with $\tau_{\text{min}}<0.94 \text{ Gyr}$, although there is a secondary mode at $\tau_{\text{min}} \approx 3$ Gyr. This secondary mode coincides with steep $\alpha \approx -2.4$, meaning there is slight support for a population dominated by relatively delayed mergers that occur rapidly once the minimum timescale is reached. This degeneracy illustrates how a lack of information at high redshifts ($z > 1.5$) allows for multiple combinations of delay time parameters to reproduce the observed low-redshift decline in the merger rate.
Additionally, there is a second mode at $\gamma \approx 4.5$, corresponding to a sharp rise in the progenitor formation rate at early times (high redshifts).

\subsection{4-point Fit}\label{sec:results_4point}

The 4-point approach considers intermediate redshift points in addition to the boundaries: $z =0$, $z=0.4$, $z=1.0$ and $z = 1.5$. 
As illustrated in the top panel of Figure \ref{fig:plot_4point}, the inclusion of intermediate redshift points forces the model to adhere to more of the structure in the \textsc{B-Spline} posterior curves. As such, there is a better visual agreement across redshift.
This improvement is quantified with the SSE value, $ \text{SSE}=2.98 ^{+13.86}_{-0.62}$ at 90\% credibility. Compared to the SSE value of the 2-point framework, the additional redshift points reduces the median SSE value by $\approx 4.5$ times. Thus, the consideration of additional points demonstrates a marked improvement in the model's global representation of the data compared to only 2 points. 

Under this framework, we find that while the 4-point method produces an overall lower SSE compared to the 2-point approach by anchoring the model at more spots, tension between the model and the data remains. Whereas the 2-point residuals in Fig. \ref{fig:plot_2point} were constrained to nearly zero at their anchors, the 4-point residuals in Fig. \ref{fig:plot_4point} reveal the conflict between the smooth nature of the model and the splinic nature of the BBH merger rates even at the chosen redshift points. At low redshifts, the model remains evenly constrained, with residuals at $z=0$ and $z=0.4$ of $\log_{10}(\mathcal{R}_{\rm data}/\mathcal{R}_{\rm model}) = 0.00^{+0.03}_{-0.03}$ and $0.00^{+0.08}_{-0.08}$, respectively. However, as the fit extends into the high-redshift regime, the variance increases and the model struggles to track the data-driven reconstructions. At $z=1.0$ and $z=1.5$, the residuals are $0.00^{+0.17}_{-0.07}$ and $0.00^{+0.03}_{-0.16}$. While these residuals remain consistent with zero, the increased spread and skewed residual distributions highlights a fundamental morphological mismatch. As noted in Section \ref{sec:results_2point}, many data-driven \textsc{B-Spline} curves exhibit a flattening following a peak after $z \approx 1$. Our physics-driven model cannot simultaneously accommodate these peaks and the surrounding slope and is forced to compromise between the anchors. Consequently, while the 4-point method provides a better global fit (lower SSE) than the 2-point method, it exposes that the physical model is ultimately unable to capture the non-trivial structure present in the LVK population results.

The recovered parameter distributions are shown in Figure \ref{fig:cornerplot_4p}. 
The best-fit delay time distribution and progenitor formation rate parameters vary from the 2-point results (Figure \ref{fig:cornerplot_2p}), as multimodalities emerge and convergence of certain parameters can no longer be achieved while fitting for more than 2 points. These multimodalities, especially because they occur at the boundaries of the searched parameter regions, suggest that our simple model struggles to fit the additional structure in the \textsc{B-Spline} posterior curves. We verified that this railing behaviour is robust, as expanding the range of boundaries does not lead to convergence.

The delay time slope $\alpha$ tends to cluster near the lower boundary value of $-3$, corresponding to very steep delay time distributions when the inferred \textsc{B-Spline} merger-rate curve rises rapidly and exceeds $200 \mathrm{Gpc}^{-3} \mathrm{yr}^{-1}$ at $z\gtrsim1$. Thus, these solutions are typically accompanied by large values of $\gamma$. Physically, steep delay time distributions concentrate mergers close to the time of binary formation, meaning that the merger rate evolution closely traces the shape of the progenitor formation history itself. As a result, a rapidly rising formation rate can efficiently reproduce the high-redshift merger rate excess of this subpopulation of \textsc{B-Spline} curves, without requiring a large overall normalization $\mathcal{A}$. Conversely, when the inferred merger rate evolution is shallower, reaching only $\lesssim100 \mathrm{Gpc}^{-3} \mathrm{yr}^{-1}$ at $z\gtrsim1$, the preferred $\alpha$ values shift toward the upper boundary, corresponding to approximately flat delay time distributions. In this case, long minimum delay times $\tau_\text{min}$ are generally disfavored, since the combination of a shallow distribution and large minimum delay would overly suppress mergers.

Additionally, the parameters $\log\mathcal{A}$ and $\delta$ display a strong degeneracy. Larger normalization values of $\mathcal{A}$ correlate with more negative values of $\delta$, corresponding to a steeper late time decline in the progenitor formation rate. A more negative $\delta$ reduces the formation rate at late times, requiring the overall normalization to be higher to account for the sudden drop. The larger values of $\log\mathcal{A}$ also correspond to higher $\tau_\text{min}$. Longer minimum delay times suppress mergers at high redshift by delaying binaries to later cosmic times, requiring the larger values of $\log\mathcal{A}$ in order to sustain the observed merger rate amplitude. Similarly, when $\delta$ is strongly negative, the fit compensates through larger values of $\gamma$, producing a sharply peaked formation history with both a rapid early time rise and rapid late time fall.

These behaviours are primarily driven by the inability of the simple parametric model to reproduce the flattening of the \textsc{B-Spline} merger rate evolution around $z\sim1$--$1.5$. Unlike the 2-point fit, the 4-point fit takes into account whether or not the merger rate peaks and turns over within the observed redshift range which can have significant impact on the permitted parameter space. {The persistent residuals in the 4-point framework can be interpreted as a manifestation of model bias: while the additional anchoring points reduce variance in the inferred merger rate reconstruction, they simultaneously expose the inability of the high-bias phenomenological model to reproduce the nontrivial structure present in the \textsc{B-Spline} posterior.}

In Fig.~\ref{fig:cornerplot_2p} and Fig.~\ref{fig:cornerplot_4p}, we find that $\delta$, the parameter governing the decline in the progenitor formation rate, is consistent between both methods. This indicates that the low redshift behavior of the progenitor formation rate is robustly constrained by the data, independent of the number of redshift anchors used in the fit.
At the same time, the 4-point framework substantially improves the reconstruction of the merger rate evolution across redshift, reducing both the typical deviation from the \textsc{B-Spline} result and the occurrence of large mismatches at non-anchored redshift points.
Motivated by this improved fidelity, we adopt the 4-point fit for further analysis of the progenitor formation rate, as it provides a more reliable reconstruction of its redshift evolution. We leverage the well-constrained parameter $\delta$ to quantify the decline of the progenitor formation rate relative to the global SFR.

\subsection{Inferred Progenitor Formation Rate}\label{sec:results_progform}

\begin{figure}
    \centering
    {\includegraphics[width=9cm]{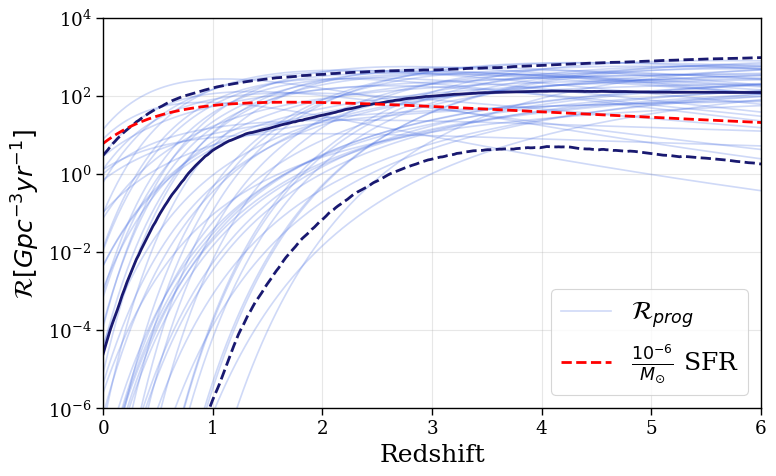} }%
   {\includegraphics[width=9cm]{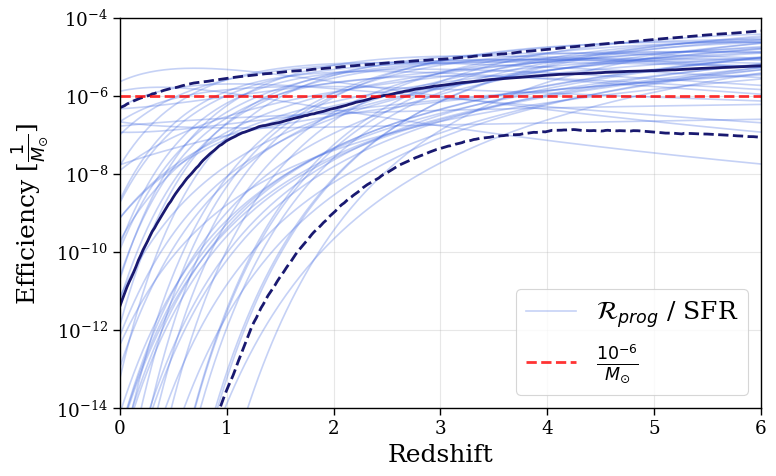} }%
    \caption{\textit{Top:} Formation rates as a function of redshift. Samples of physics-driven model inferred progenitor formation rates (blue) with the median denoted by solid dark line, and $5^{th}$ and $95^{th}$ percentile bands denoted by dark dashed lines. This is compared to the global SFR, scaled down (red) \citep{Katsianis}. \textit{Bottom:} Efficiency of BBH progenitor formation rates as a function of redshift (blue). The $5^{th}$ and $95^{th}$ percentile bands are denoted by dark pointed lines. The red dashed line indicates the rescaled SFR.}
    \label{fig:progenitorplot}
\end{figure}

The progenitor formation rates we infer from our 4-point fit are shown in Figure \ref{fig:progenitorplot} (top panel). We observe the GWTC-4.0 inferred progenitor rates peak at earlier times (higher redshifts) than the global SFR and declines more rapidly towards late times (lower redshifts) than the SFR.
This is also apparent in the recovered $\delta$ parameter, which is more negative for BBH progenitors than for the global SFR (represented with the red dashed line in Fig.\ref{fig:cornerplot_4p}), indicating the decline in the BBH progenitor rate is steeper than that of the SFR. We compute the global SFR and inferred progenitor rate slope values between redshift $z_1=0.1$ and $z_2=1.0$ to quantify the variation in the slope steepness. The SFR's slope (in log-space) is $\frac{\log_{10}(\mathcal{R}_1) - \log_{10}(\mathcal{R}_2)}{z_1 - z_2} \approx 0.86$. 
On the other hand, the log-space slope of the inferred progenitor formation rate (in Figure \ref{fig:progenitorplot}) is $\frac{\log_{10}(\mathcal{R}_1) - \log_{10}(\mathcal{R}_2)}{z_1 - z_2}=4.58^{+15.39}_{-0.97}$ at 90\% credibility. The median slope of the progenitor formation rate is approximately 5.3 times larger than the SFR's. 

We divide the inferred progenitor formation rate by the SFR to obtain the fraction of stellar mass that goes on to participate in a BBH merger, or the efficiency, as a function of $z$ (Eq.~\ref{eq:efficiency}). This is shown in the bottom panel of Figure \ref{fig:progenitorplot}, where we see a decline in the efficiency as cosmic time advances. 
The behavior of the inferred efficiency as a function of redshift is expected since BBH formation preferentially occurs in earlier, more metal poor times, and declines steeply towards late times as the Universe becomes more enriched. 
This is consistent with theoretical predictions, as well as observational results from GWTC-3.0 (e.g. \citep{Fishbach_vonson2023}, \citep{schiebelbeinzwack2024massdensitymergingbinary}).

\section{Conclusion}\label{sec:conclusion}

{In this work, we introduce a curve-fitting framework that performs an interpretable projection of nonparametric posteriors, mapping data-driven reconstructions of the BBH merger rate evolution onto physically meaningful formation parameters. This enables residuals and SSE values to directly expose model bias and incompatibilities between flexible data-driven reconstructions and phenomenological descriptions.}

We fit a simple model that parametrizes the BBH merger rate with a progenitor formation rate (Eq. \ref{eq:prog}) and delay time distribution (Eq. \ref{eq:ptau}).
Rather than fitting this model directly to the gravitational wave data using hierarchical Bayesian inference, we applied a curve-fitting method to an existing nonparametric fit to GWTC-4.0 data: the \textsc{B-Spline} fit. For each \textsc{B-Spline} posterior draw, we find the best-fitting progenitor formation rate and delay time distribution parameters. 

A benefit of our methodology is the ease of including an arbitrary number of redshift data points to anchor the physics-driven model. By increasing the number of redshift points used to infer the BBH formation parameters from two to four, we force the model to adhere to more structure of the BBH merger rate.
Our 4-point curve fitting results capture the overall shape of the observed BBH merger rate evolution (Fig. \ref{fig:plot_4point}) and shows an improvement in matching the data compared to the 2-point fit, as supported by the decrease in SSEs (Eq.~\ref{eq:SSE}) by a factor of $\approx 4.5$. However, the 4-point fit still fails to capture all of the structure found by the \textsc{B-Spline} posterior as shown by the residuals in Fig. \ref{fig:plot_4point}. This causes the distribution of the best-fit parameters to often cluster at the boundaries of the searched region (Fig. \ref{fig:cornerplot_4p}). Certain \textsc{B-Spline} posterior draws exhibit erratic changes in slope between redshifts $z = 0$, 0.4, 1.0 and 1.5, which our simple form for the progenitor formation rate and delay time distribution struggles to reproduce. 

However, the ability to investigate the residuals between a given parametric fit and the data-driven fit is a powerful advantage of our method. In standard hierarchical Bayesian population inference, the model typically latches onto the subset of posterior draws it can describe well, effectively ignoring data-driven curves that are incompatible with the chosen functional form. This can mask fundamental model misspecification.
By evaluating the residuals and SSE for every draw in the \textsc{B-Spline} posterior, we gain a transparent view of where the parametric model fails to replicate the data.

We analyze the inferred progenitor formation rates from the 4-point fit, motivated by the convergence of the decline in the progenitor formation rate, $\delta$. In agreement with past work, we find that the progenitor formation rate peaks earlier, and declines more rapidly at late times, than the global SFR (Figure~\ref{fig:progenitorplot} (bottom panel)). In other words, the BBH progenitor formation efficiency drops off rapidly at late times (low redshifts; Figure~\ref{fig:progenitorplot} (bottom panel)). 
This is quantified by the median progenitor formation rate slope being $\approx 5.3$ times steeper than the global SFR's slope between $z=0.1$ and z=$1.0$.  
This steeper slope is consistent with theoretical predictions that the BBH progenitor formation is more efficient in low-metallicity environments, which are more prevalent in earlier cosmic epochs. 

Although we have applied our least squares curve-fitting method to fit a simple parametric form for the merger rate evolution in this work, future studies could apply our method to fit physics-driven models to gravitational wave data without having to evaluate likelihoods. This would allow us to easily and inexpensively find best-fit physical parameters (e.g., mixture fractions between different formation channels, the parameters governing stellar environments, binary stellar evolution, and/or dynamics). Furthermore, it would allow us to investigate the residuals between the physics-driven and the data-driven models that may point to missing physics or modeling systematics. This framework can be a useful tool in extracting more precise insights into the factors affecting BBH formation and evolution. 

\begin{acknowledgments}
We thank Ivan Juarez-Reyes for comments on an early draft of our manuscript, as well as Amanda Farah, Ben Farr, and Daniel Holz for helpful discussions.
ASZ acknowledges the support of the Natural Sciences and Engineering Research Council of Canada - Canada Graduate Scholarships - Doctoral (NSERC-CGS-D) program. 
EB, ASZ, and MF acknowledge support from the Natural Sciences and Engineering Research Council of Canada under grant RGPIN-2023-05511, the Alfred P. Sloan Foundation, the University of Toronto Connaught Fund, and the Ontario Early Researcher Award ER24-18-170.
This material is based upon work supported by NSF’s
LIGO Laboratory which is a major facility fully funded
by the National Science Foundation.
\end{acknowledgments}

\bibliography{main}
\bibliographystyle{aasjournalv7}

\end{document}